# A Neutron Scattering Perspective on the Structure, Softness and Dynamics of the Ligand Shell of PbS Nanocrystals in Solution

T. Seydel[1], M. M. Koza[1], O. Matsarskaia[1], A. André[2], S. Maiti[3,#], M. Weber[2], R. Schweins[1], S. Prévost[4*], F. Schreiber[3,5], M. Scheele[2,5]

[1]Institut Max von Laue - Paul Langevin (ILL), 71 Avenue des Martyrs, CS 20156, 38042 Grenoble Cedex 9, France

[2]Institute of Physical and Theoretical Chemistry, University of Tübingen, Auf der Morgenstelle 18, 72076 Tübingen, Germany.

[3]Institute of Applied Physics, University of Tübingen, Auf der Morgenstelle 10, 72076 Tübingen, Germany

[4] ESRF-- The European Synchrotron, 71 Avenue des Martyrs, CS 40220, 38043 Grenoble Cedex 9, France

[5]Center for Light-Matter Interaction, Sensors & Analytics LISA+, University of Tübingen, Auf der Morgenstelle 15, 72076 Tübingen, Germany.

*present address: Institut Max von Laue - Paul Langevin (ILL), CS20156, 38042 Grenoble, France

# Present address: Institute of Neuroscience and Medicine, Forschungszentrum Jülich, 52425 Jülich, Germany

**Abstract.** Small-angle neutron and X-ray scattering, neutron backscattering and neutron time-of-flight spectroscopy are applied to reveal the structure of the ligand shell, the temperature-dependent diffusion properties and the phonon spectrum of PbS nanocrystals functionalized with oleic acid in deuterated hexane. The nanocrystals decorated with oleic acid as well as the desorbed ligand molecules exhibit simple Brownian diffusion with a Stokes-Einstein temperature-dependence and inhibited freezing. Ligand molecules desorbed from the surface show strong spatial confinement. The phonon spectrum of oleic acid adsorbed to the nanocrystal surface exhibits hybrid modes with a predominant Pb-character. Low-energy surface modes of the NCs are prominent and indicate a large mechanical softness in solution. This work provides comprehensive insights into the ligand-particle interaction of colloidal nanocrystals in solution and highlights its effect on the diffusion and vibrational properties as well as their mechanical softness.

**Introduction.** Electron-phonon interactions greatly influence the optoelectronic properties of semiconductor nanocrystals (NCs).[1–7] The occurrence of a phonon bottle-neck and the associated feasibility of multi-exciton generation in NCs is an important consequence of their vibrational properties, which requires a detailed understanding for an exploitation in optoelectronic devices.[8–12] Recent investigations of the phonon spectrum of PbS NCs indicate that mechanical softness due to dangling bonds and other defects on the surface of the NCs invokes a large density of low-energy phonons with strong coupling rates to electrons.[13] It has been argued that these low-energy phonons may be responsible for the surprisingly fast electron relaxation rates in many semiconductor NCs and that their distribution is linked to the chemical nature and structure of the ligand shell.[14,15] While these studies were performed in the solid state and after ligand exchange with short cross-linkers such as 1,2-ethanedithiol, methanethiol or halide ions, no comparable knowledge exists about the phonon spectrum of PbS NCs in solution, terminated with its native ligand oleic acid (OA). In related work, the structure of the ligand shell of OA-terminated NCs has been explored to some degree in the solid state, but remains scarcely studied in solution.[16–20]

Any understanding of the structure of the ligand shell in solution requires an account for dynamic phenomena, possibly different forms of diffusion, ligand desorption and phase changes of the ligand shell, for instance due to a change in temperature.[21–24] Previous studies of the diffusion of NCs in solution by electron microscopy have revealed orders of magnitude smaller diffusion coefficients ($D$) than predicted by the Stokes-Einstein equation.[25–30] In contrast, diffusion ordered nuclear magnetic resonance spectroscopy (DOSY) of NC solutions yielded $D$-values in good agreement with Stokes-Einstein.[21,31–34] Molecular dynamics simulations have introduced several different models for the structure of a typical NC ligand shell in the solid state, ranging from fully erected ligand molecules ("spiky ball" model) over bundles of unfolded ligands ("coiled spiky ball") to a layer of collapsed ligands ("wet hair" model).[17] For relatively bulky molecules like OA, the wet hair configuration has been identified as the most stable structure. In contrast, small angle neutron scattering (SANS) on metal NCs implied that a fully unfolded ligand structure prevails in solution.[18,35]

In this context, a number of fundamental questions arise. Is the structure of the ligand shell in solution identical to that in the solid state? How confined are ligands bound to the surface? Is the mechanical softness of the NCs, that is a large density of low energy phonons, also present in solution? How do these properties vary with temperature, especially close to phase transitions of the ligand or the solvent? Is there evidence for a correlation between changes of the dynamics or the structure with variations of the phonon spectrum?

We address these questions for PbS NCs with a radius of 3.3 nm, terminated with OA in $d_{14}$-hexane by applying a combination of several neutron scattering and spectroscopic techniques. With SANS and small angle X-ray scattering (SAXS), we determine the width of the ligand sphere. By quasi-elastic neutron scattering (QENS), we derive the temperature-dependent diffusion coefficients of the NCs, distinguish between the dynamics of adsorbed and desorbed ligands, derive the mean-free-path of the bound ligands and determine phase transitions between 183 K – 295 K. We use inelastic neutron spectroscopy (INS) to record the generalized density of states (GDOS) between 100 K and 239 K and show for the first time that surface phonons of the NCs as well as vibrations of the ligands can be recorded in solution.

**Methodology.** While details to the experimental procedures are provided in the Supporting Information, here we briefly summarize the most important aspects of the scattering experiments performed in this work.

*Small-Angle Neutron and X-ray Scattering (SANS and SAXS).* Scattering contrast in SAXS is provided by the difference in electron density, which makes SAXS the ideal tool for determining the structure of the NC cores due to their large density of heavy Pb atoms against a background of light elements ($d_{14}$-hexane or oleic acid).[36] In contrast, the scattering cross-section for neutrons varies non-monotonically with the isotope number.[37] As detailed in the Supporting Information (**Figure S2**), the SANS signal and the resulting scattering contrast is dominated by scattering of the H-rich ligands, considering that the solvent is deuterated with 99 % isotope purity.

*Quasi-Elastic Neutron Scattering QENS.* QENS is a powerful method for studying the diffusion of organic matter in solution, which is based on energy-resolved scattering of neutrons by the H-nuclei.[38] Diffusion of the molecules during the scattering event invokes a broadening of the elastic scattering line, which allows to extract the diffusion coefficient $D$. This method can be employed for a wide range of momentum transfers $q$, at which the energy transfer $\omega$ of the neutrons to the sample is recorded. The $q$-dependence of the scattering signal contains information on the confinement geometry of the diffusive motion of the scatterers, specifically its elastic incoherent structure factor (EISF) $A_0(q)$. The EISF is the long-time limit of the intermediate scattering function and a rough estimate for the mean-free-path. Therefore, $q$-dependent QENS experiments can determine the degree of confinement of a diffusing scatterer, which is essentially impossible by other techniques.

Selecting a fixed energy transfer window allows following changes in the scattering intensity with temperature, which contains information on slowed diffusion due to freezing and possible phase changes. These scans only require a recording time on the order of one minute per temperature step and can, thus, be acquired much more quickly than full QENS spectra with a recording time of several hours per spectrum. Elastic fixed window scans (EFWS, energy transfer = 0 µeV) monitor intensity changes of scatterers which are stationary over the observed time scale of a few nanoseconds. Inelastic fixed window scans (IFWS, energy transfer ≠ 0 µeV) exclusively record intensity changes due to the dynamics of the scatterers. Thus, IFWS *vs.* EFWS experiments allow distinguishing temperature-dependent changes in the bulk diffusion properties between fast and slow scatterers, such as free OA *vs.* OA bound to the surface of NCs.

*Inelastic Neutron Scattering (INS) by Time-of-Flight Spectroscopy.* A time-of-flight (TOF) experiment employs the velocity change of neutrons during scattering from a sample to calculate the energy exchange. The high-energy limit for TOF spectroscopy is much larger (>48 meV) than for the backscattering experiment employed during QENS, which includes not only the energy range of solvent diffusion (e.g. $d_{14}$-hexane) but also that of vibrational excitations. Basically, vibrations are related to the chemical bonding and their frequencies are a measure of the bond strength. The distribution of vibrational excitations contains information on the chemical interaction between the NCs and the adsorbed OA molecules. In case of a strong interaction, we expect the formation of hybrid vibrational modes, procuring a renormalization of the bulk vibrational spectra of the individual systems. As the contribution of Pb and S to the recorded signal is negligible according to its weak relative neutron scattering power in the specimen (see **Table S3** and **Figure S2** in SI for details), INS monitors the vibration of OA with high sensitivity. The inelastic response and its changes are provided in the form of the spectral distribution to which we refer to as the generalized density of states (GDOS).

**Results.** Oleic acid (OA)-capped PbS NCs are dissolved in $d_{14}$-hexane to yield a stable colloidal solution with a concentration of 140 µmol/L, corresponding to a volume fraction of 1.6 % (for details, see Supporting Information). SANS and SAXS of PbS/OA/d$_{14}$-hexane are shown in **Figure 1** as functions of the momentum transfer $q = |\vec{q}| = \frac{4\pi}{\lambda}\sin\theta$, where $2\theta$ is the scattering angle and $\lambda$ is the incoming wavelength.

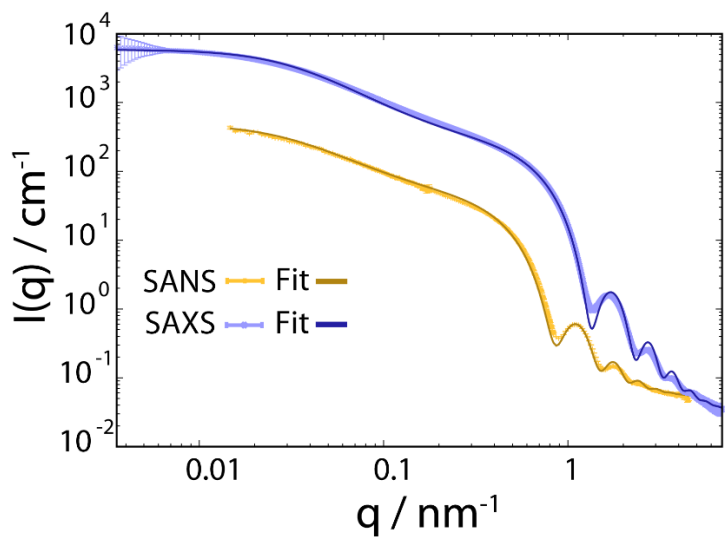

**Figure 1**. Small angle neutron and X-ray scattering (SANS and SAXS) spectra of PbS NCs decorated with oleic acid in d$_{14}$-hexane at 298 K. The bold lines are fits to the data and are described in detail in the Supporting Information.

The different onsets of the oscillations in the SANS and SAXS data starting at $q \approx 1$ nm$^{-1}$ reflect the different sensitivities of the two techniques to the PbS core and the OA shell as explained in the methodology section. From the SAXS data, we calculate the radius of the PbS core as 33 Å and the thickness of the OA shell as 18 Å from SANS. The latter value roughly corresponds to one molecular length of OA (19.5 Å; from universal force field calculations). It is also consistent with a SANS/SAXS study on oleylamine-capped Au NCs in toluene, which reported 17 Å for the width of this similar ligand sphere.[18]

We model the SANS and SAXS data in **Figure 1** with core-shell spheres, a fraction of which is included in mass fractal aggregates (For details, see SI). Both datasets display a large excess of scattering intensity at low $q$ compared to the scattering expected from well-dispersed core-shell spheres. The SAXS data indicates the exclusive presence of elongated aggregates as evidenced by the power-law decay of the intensity at mid $q$ that follows $I_{SAXS}(q) \approx q^{-1}$, as typical for rod-like scatterers with a fractal dimension close to 1. The SANS data, however, are different and allow to discard the hypothesis that all particles are part of rod-like clusters. By simultaneously fitting the SANS and SAXS data, we establish that there are two populations in the PbS/OA/$d_{14}$-hexane sample: 1) a large majority of spherical monomers corresponding to well-dispersed NCs and 2) a smaller fraction (21 %) of NC aggregates. We calculate the OA/PbS ratio in the aggregates to be ~30% smaller than that in the well-dispersed NCs, indicating that partial ligand deficiency may be responsible for this finding.

**Figure 2** displays the QENS signal recorded on IN16B at $q$ = 0.29, 0.44 and 0.82 Å$^{-1}$, and $T$ = 239 K, of OA-decorated PbS NCs in $d_{14}$-hexane (left column, **orange circles**). For comparison, we also measure solutions of 35 mmol/L OA in $d_{14}$-hexane without PbS NCs (right column, **blue circles**), as well as of pure $d_{14}$-hexane in the same conditions (**black squares**). The concentration of free OA in the OA/$d_{14}$-hexane reference sample is chosen to match the nominal concentration of bound OA on the NC surface in the PbS/OA/$d_{14}$-hexane sample measurement (for details, see Supporting information). The significant broadening of the QENS peak visible in the PbS/OA/$d_{14}$-hexane spectrum (**Figure 2**, left column) indicates diffusion on a nanosecond time scale. A narrower QENS peak is also visible for the OA/$d_{14}$-hexane sample (**Figure 2**, right column).

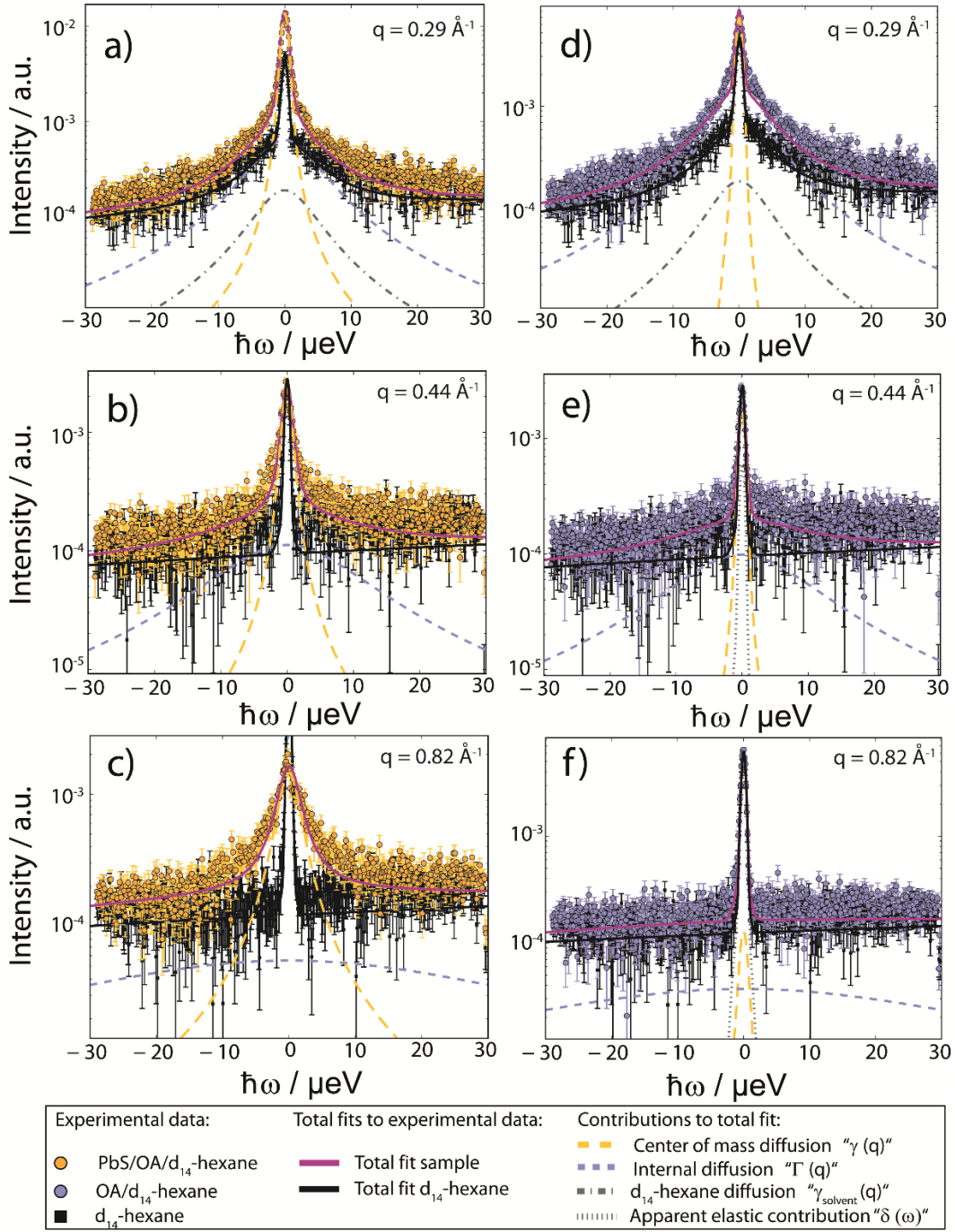

**Figure 2**. QENS spectra at T = 239 K and three different scattering vectors. **a-c)** Experimental data of PbS/OA/$d_{14}$-hexane and **d-f)** of OA/$d_{14}$-hexane, including a $d_{14}$-hexane background in all cases. Solid lines represent fits to the experimental data, composed of the Lorentzian contributions from the center-of-mass diffusion, internal diffusion, $d_{14}$-hexane solvent and an apparent elastic contribution represented by dashed lines according to the model described by equation (1) in the Supporting information. The apparent elastic contribution and the $d_{14}$-hexane diffusion are not visible on all plots due to the low signal. The $d_{14}$-hexane contribution has been rescaled by the effective excluded volume of OA. All data has been binned by pairs of two channels subsequent to fitting for better visibility.

While details are provided in the Supporting information, we fit the QENS spectra according to a heuristic model, which assumes a convolution of three Lorentzian contributions due to 1) a center-of-mass diffusion of the NCs (or alternatively of OA vesicles, cf. further below), 2) internal diffusion of OA molecules moving along the NC surface and 3) a solvent diffusion background.[39,40] The three Lorentzian contributions give rise to a line broadening, characterized by the widths $\gamma(q)$, $\Gamma(q)$ and $\gamma_{solvent}(q)$, respectively. The contribution by the center-of-mass diffusion scales with the scalar fit parameter $A_0(q)$, which is the Elastic Incoherent Structure Factor (EISF) associated with the spatially confined diffusion of OA. In addition, we include a delta-function $\delta(\omega)$ due to an apparent elastic contribution, e.g. by the empty can and/or very large objects that appear essentially immobile on the time scale accessible to IN16B (e.g. the agglomerates observed in SAXS, **Figure 1**). These four contributions are displayed in **Figure 2** as dashed lines. Their convolution according to the heuristic model is depicted as solid lines for PbS/OA/$d_{14}$-hexane and OA/$d_{14}$-hexane (purple) as well as for pure $d_{14}$-hexane (black). The excellent fit to the experimental QENS data supports the applicability of the model.

We now use the Lorentzian widths, $\gamma(q)$ and $\Gamma(q)$, to extract the corresponding scalar diffusion coefficients for the center-of-mass diffusion ($D_1$) and internal diffusion ($D_2$). We assume Fickian diffusion and impose $\gamma(q) = D_1 q^2$. Similarly, we test different assumptions for $\Gamma(q)$, namely the so-called jump-diffusion $\Gamma(q) = \frac{D_2 q^2}{1 + D_2 q^2 \tau}$, and Fickian diffusion as $\Gamma(q) = D_2$ $q^2$. When employing the jump diffusion model, we obtain the best fit by setting the free scalar parameter $\tau$ to zero, which reduces the model again to simple Fickian diffusion. Therefore, we subsequently impose $\Gamma(q) = D_2 q^2$, which is in accordance with earlier reports for the diffusion of lipid molecules within lipid vesicles.[41] (We note that, due to the limited observable time scales of motions accessible on IN16B, $D_2$ does not necessarily account for all types of molecular diffusive motions that OA may be subject to). The results for the extracted diffusion coefficients are displayed in **Figure 3** for three different temperatures $T$ = 183 K, 239 K and 295 K. We find that the experimental $D_1$ values for the PbS/OA/d$_{14}$-hexane sample (**Figure 3**, **orange squares**) are close to the expected bulk diffusion of spherical objects according to the Stokes-Einstein equation. To illustrate this, we calculate the Stokes-Einstein diffusion coefficient using $r$ = 5.1 nm (from **Figure 1**) and $\eta$ = 1.84, 0.58 and 0.31 mPa*s as the dynamic viscosity of hexane at $T$ = 183, 239 and 295 K, respectively (**Figure 3**, **purple triangles**). For the OA/d$_{14}$-hexane sample, our fit gives very small values for the center-of-mass diffusion constant, here termed $D_1{'}$ **(orange circles)**. We speculate that these small values arise from nearly immobile OA vesicles suspended in d$_{14}$-hexane. In this picture, which will be corroborated by the discussion of the fixed-window data in the following section, large OA vesicles form only in the absence of the PbS nanoparticles, whilst the OA decorates the PbS in their presence.

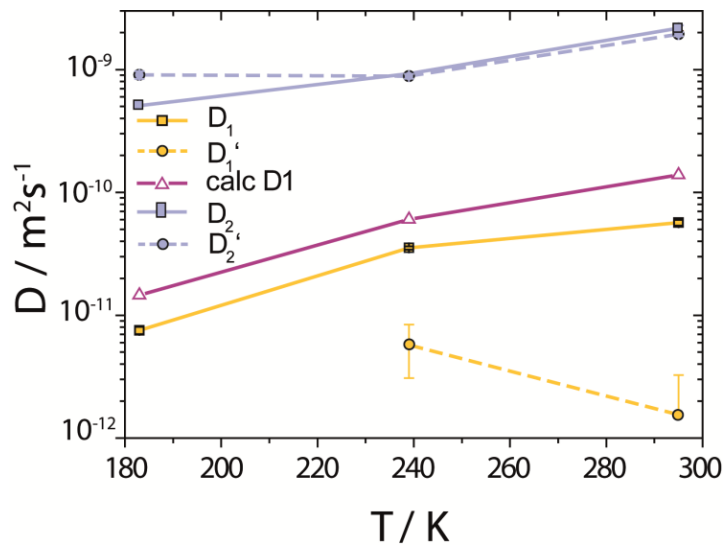

**Figure 3.** Diffusion coefficients in $d_{14}$-hexane extracted from the QENS fit in Figure 2. Center-of-mass diffusion of PbS/OA NCs ($D_1$) or OA vesicles ($D_1'$) in orange; internal diffusion of OA on the NC surface ($D_2$) or within OA vesicles ($D_2'$) in blue. Calculated diffusion coefficient for the PbS/OA NCs according to the Stokes-Einstein equation in purple.

The evaluation of the internal diffusion contribution to the QENS fits yields similar $D_2$ values for both, the PbS/OA/$d_{14}$-hexane ($D_2$, **blue squares**) and OA/$d_{14}$-hexane ($D_2'$, **blue circles**) samples in **Figure 3**. We speculate that $D_2$ and $D_2'$ are associated with the diffusion of OA along the surfaces of the PbS NCs (PbS/OA/$d_{14}$-hexane sample) and within the OA vesicle membranes (OA/$d_{14}$-hexane sample), respectively. This speculative picture would describe two-dimensional Fickian diffusion within a surface. The picture is supported by the corresponding hydrodynamic radii calculated by the Stokes-Einstein equation summarized in **Table 1** in the presence of PbS ($r_2$) and for OA/ $d_{14}$-hexane only ($r_2'$). The nominal radii of 3.2 Å and 3.6 Å are in good agreement with the previously reported value of 3.1 Å for similar fatty acids in hexane at 298 K.[42]

|  | 183 K | 239 K | 295 K |
|---|---|---|---|
| $r_2$ (PbS/OA/$d_{14}$-hexane) | 1.4 Å | 3.3 Å | 3.2 Å |
| $r_2'$ (OA/$d_{14}$-hexane) | 0.8 Å | 3.4 Å | 3.6 Å |
| r(OA/hexane), ref. 42 |  |  | 3.1 Å |

**Table 1.** Calculated hydrodynamic radii for OA according to the diffusion coefficients $D_2$ and $D_2'$ in Figure 3 and the Stokes-Einstein equation. $r_2$ represents the radii in the presence of the NCs, while $r_2'$ stands for the calculated radii without the NCs. For comparison, we note the literature value for the radius of similar fatty acids reported in ref.42.

To further investigate the spatial confinement of the internal OA diffusion, we utilize the EISF $A_0(q)$ obtained by fitting the QENS data, which is depicted in **Figure 4** for the PbS/OA/$d_{14}$-hexane sample. The EISF data is fit with a heuristic model described in ref. 43 and 44, which assumes a superposition of confined diffusion with a characteristic mean-free-path $d$ and molecular reorientation jumps. From the onset of the rise in the EISF data towards $q = 0$, we gauge the mean-free-path as $r = 2\pi/\text{onset} = 9$ Å at 295 K. The corresponding fit of the data is obtained with a mean value for $r$ of 6.6 ± 1.9 Å. The reasonable agreement between these two independent estimates supports the applicability of the model. We note that the significant standard deviation in the $A_0(q)$ prevents a more involved modeling and correction for the Debye-Waller factor.

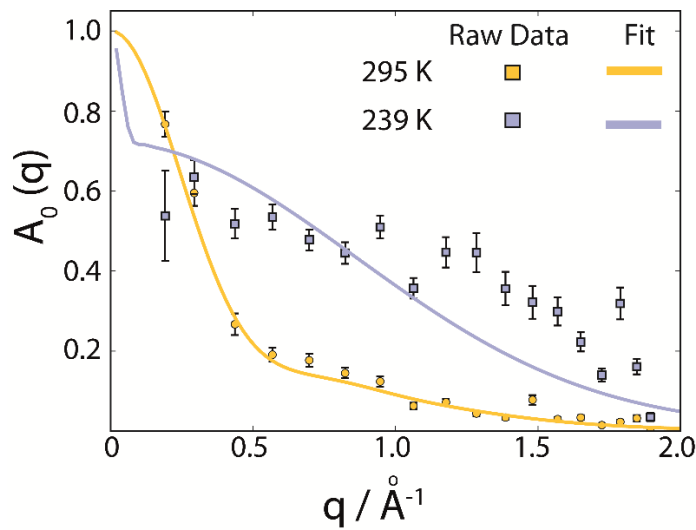

**Figure 4**. $A_0(q)$ at 239 K and 295 K for PbS/OA in $d_{14}$-hexane obtained from the fit of the QENS spectra (symbols). Solid lines represent qualitative fits according to the heuristic model given in ref. 43 and ref. 44 under the assumption of diffusion in a sphere with a smeared-out radius $R$ of 6.6 ± 1.9 Å Å for T=295 K, including a background of immobile scatterers.

At 239 K, no stable fit of $A_0(q)$ can be obtained, presumably due to the growing contribution of large NC agglomerates (cf. the SAXS data in **Figure 1**). The generally higher $A_0(q)$ values at 239 K compared to 295 K indicate that a smaller fraction of OA is mobile at lower temperature. We note that at 183 K, $A_0(q)$ is mostly flat and no stable fit may be obtained, which is why we omit it here. We argue that at this low temperature, the OA molecules are fully immobilized, which prevents a meaningful analysis of $A_0(q)$. Furthermore, a similar analysis applied to the OA/$d_{14}$-hexane data does not result in a discernable onset for any of the three temperatures. This suggests that the diffusion of OA is only locally confined in the presence of the NCs.

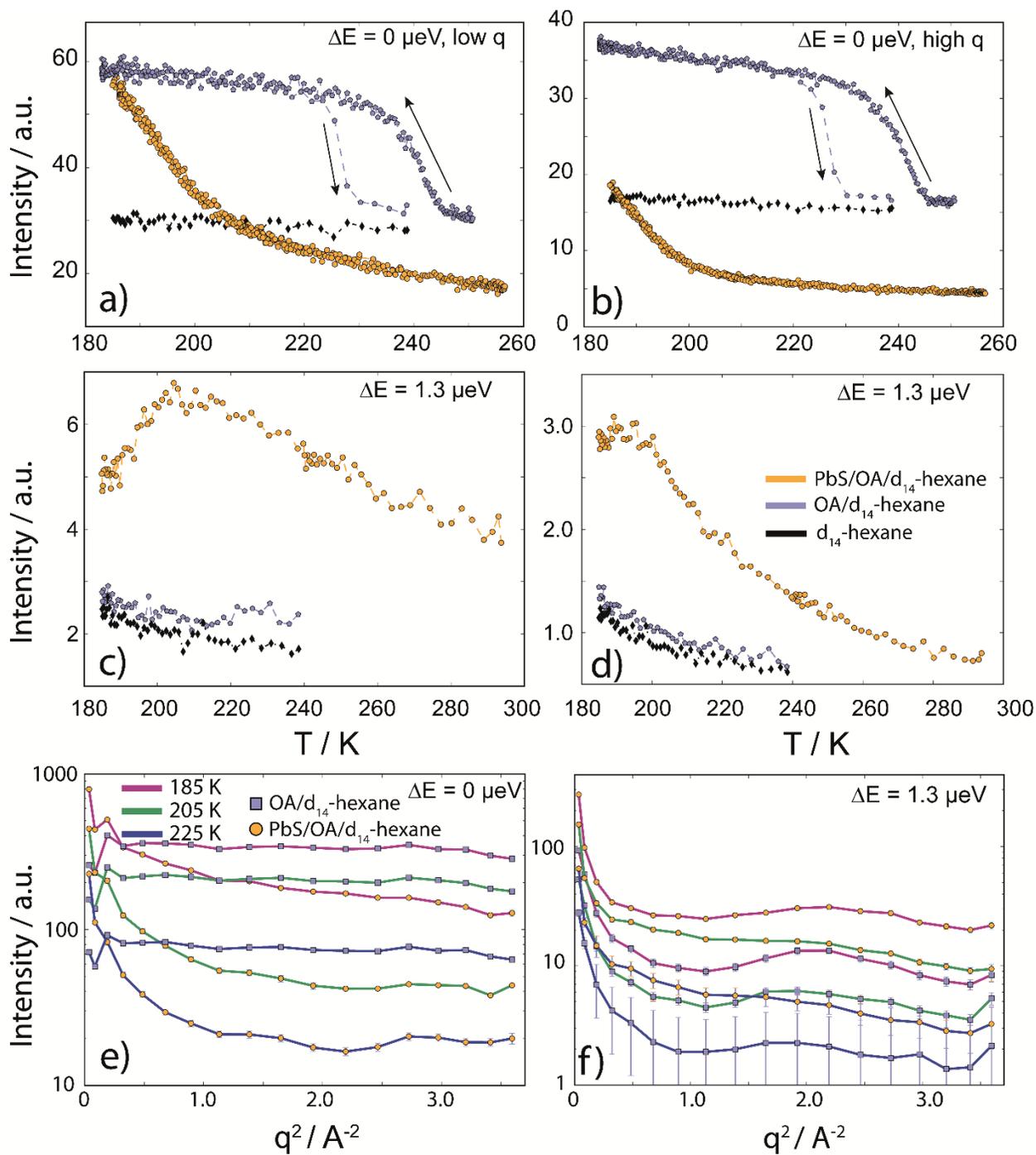

**Figure 5 a)** and **b)** Elastic fixed window scans for q = 0.44 – 0.82 Å$^{-1}$ and q = 1.39 – 1.65 Å$^{-1}$, respectively. The elastic data have been recorded both in heating and in cooling, showing a hysteresis for the OA/d14-hexane sample (indicated by black arrows). **c), d)** Inelastic fixed window scans with an offset of ΔE = 1.3 μeV for both q-ranges, respectively. The color code is identical for all panels: PbS/OA/$d_{14}$-hexane in **orange**, OA/$d_{14}$-hexane in **purple** and $d_{14}$-hexane is shown in **black**. The intensity scale is in arbitrary, yet identical units, which allows for a direct comparison at the same q-range and energy offset. *q*-Dependent intensities at **e)** 0 µeV and **f)** 1.3 µeV for PbS/OA/$d_{14}$-hexane (orange circles) and OA/$d_{14}$-hexane (blue squares) at T = 185±5 K (purple lines), 205±5 K (green lines) and 225±5 K (blue lines).

In **Figure 5**, we display EFWS (**Figure 5a,b**) and IFWS (**Figure 5c,d**) for the fixed *q*-ranges 0.44 – 0.82 Å$^{-1}$ and 1.39 – 1.65 Å$^{-1}$ of PbS/OA/$d_{14}$-hexane, OA/$d_{14}$-hexane and $d_{14}$-hexane versus the sample temperature. The OA/$d_{14}$-hexane mixture evidences a freezing step between 230 – 220 K which may be associated with immobilized OA, e.g. due to the formation of OA clusters (**Figure 5a,b; blue circles)**. As is typical for freezing transitions, the melting during heating occurs at a higher temperature than the freezing during cooling, as evidenced by the hysteresis (about 10 – 15 K in this case). Note that the freezing point of pure OA is significantly above this temperature window (278 K). For a discussion of phase transitions in these systems and the comparison bulk vs surface (flat) vs surface (curved), see references 45, 46 and therein. There is no significant QENS broadening on the ns scale even at 239 K for this mixture, consistent with the absence of a strong signal or changes in the IFWS (**Figure 5 c,d; blue circles**). The absence of Bragg scattering below the freezing step at 230 – 220 K suggests that the elastic scattering arises solely from incoherent scattering of OA and not from $d_{14}$-hexane that remains liquid (cf. also **Figure 5 e,f**). In line with this observation, pure $d_{14}$-hexane shows no significant change in the inspected temperature range in EFWS scans (**Figure 5a,b; black circles**) and only very minute changes in the IFWS (**Figure 5c,d; black circles**). We note that the diffusion of hexane slows down significantly towards the lower end of our temperature window ($6.2*10^{-10}$ $m^2s^{-1}$ at 193 K), which is in the detectable range of the QENS experiment performed here.

In the presence of PbS NCs, we observe no freezing step in the elastic signal (**Figure 5a,b; orange circles**). Instead, the elastic intensity gradually increases with decreasing temperature. Moreover, the IFWS exhibits a maximum in its temperature-dependence, at least for some values of $q$ (**Figure 5c,d; orange circles**). These observations are consistent with the strong nanosecond QENS signal seen at lower temperature.

As depicted in **Figure 5e,f**, the fixed window scans can be further illustrated by slices along $q$ for the three temperatures 225 K (**blue lines**), 205 K (**green lines**) and 185 K (**purple lines**), comparing PbS/OA/$d_{14}$-hexane (**orange circles**) and OA/$d_{14}$-hexane (**blue squares**), respectively. The elastic signal (**Figure 5e**) resembles the behavior of a harmonic solid for the OA/$d_{14}$-hexane sample, as evidenced by the straight lines in the logarithmic plot versus $q^2$. In contrast, the $q^2$-dependence of the signal from PbS/OA/$d_{14}$-hexane strongly deviates from a straight line in this plot, supporting the presence of diffusive behavior. The inelastic signal in **Figure 5f** illustrates the overall much weaker intensity of OA/$d_{14}$-hexane compared to PbS/OA/$d_{14}$-hexane, which is in line with the interpretation of the elastic signal.

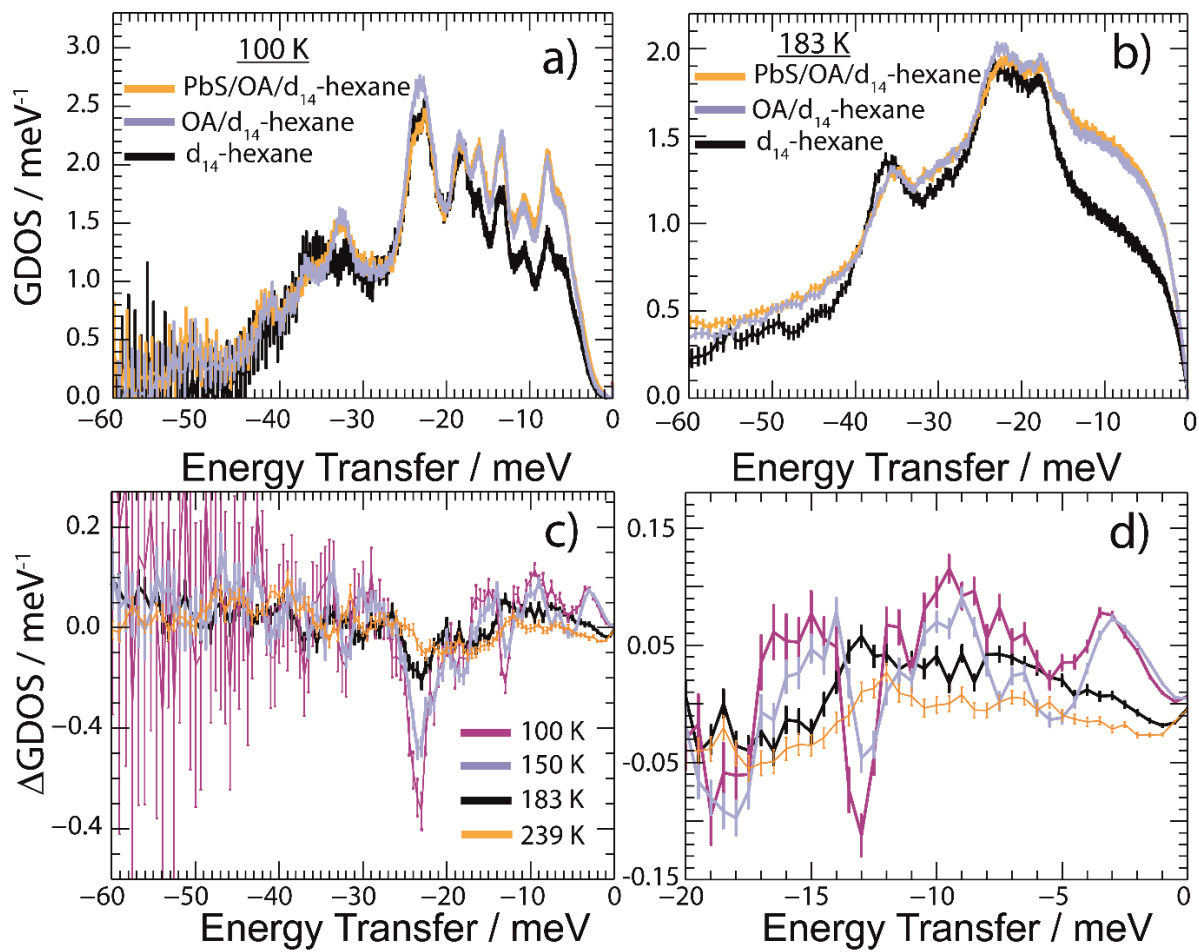

**Figure 6.** GDOS of PbS/OA/$d_{14}$-hexane, OA/$d_{14}$-hexane and pure $d_{14}$-hexane **a)** below (100 K) and **b)** above (183 K) the melting point of $d_{14}$-hexane. The $d_{14}$-hexane data have been rescaled to account for the absent signal from OA in comparison to OA/$d_{14}$-hexane and PbS/OA/$d_{14}$-hexane. **c)** Difference GDOS spectra of PbS/OA/$d_{14}$-hexane after subtraction of the OA/$d_{14}$-hexane at different temperatures. **d)** Zoom into the low-energy region of the same ΔGDOS spectra.

We report in **Figure 6** the vibrational properties of the PbS/OA/$d_{14}$-hexane, OA/$d_{14}$-hexane and $d_{14}$-hexane specimens. For convenience, the GDOS of PbS/OA/$d_{14}$-hexane and OA/$d_{14}$-hexane displayed in **Fig. 6a** at 100 K and **Fig. 6b** at 183 K are normalized to 60 vibrational modes corresponding to the vibrational degrees of freedom of a hexane molecule. The GDOS of $d_{14}$-hexane is corrected for the absent OA contribution. In the solid state at 100 K, the frozen solvent $d_{14}$-hexane dominates the spectral density in any of the specimens evidenced by the number and position of distinguishable peaks. They are well defined indicating their long lifetimes. This property changes visibly above the melting point of $d_{14}$-hexane (**Figure 6b**). Note that all GDOS preserve their relative intensities, however, they are strongly altered towards rather featureless profiles. As expected, the liquid state of $d_{14}$-hexane opens up effective relaxation channels for the vibrational modes. As a consequence, the inter-molecular bonds are short-lived leading to the enhanced broadening of their characteristic vibrational peaks and a complete loss of definition in the experimental spectra. This is less the case for the localized intra-molecular vibrations. Thus, we may associate the remaining characteristic maxima in the GDOS at 183 K with some intra-molecular properties determined by the C-H and C-D bonds. We may further conclude that the additional sharp peaks obvious in the frozen state are determined by the inter-molecular interactions.

Comparing the GDOS of PbS/OA/$d_{14}$-hexane and OA/$d_{14}$-hexane, we observe that the presence of the PbS NCs results in an additional, however weak, intensity gain below 12 meV and a visible intensity loss up to the peak of highest intensity at 20-28 meV. Thus, there is a renormalization of vibrational modes of the OA/$d_{14}$-hexane system in the presence of the PbS NCs. Note that the energy range of up to 28 meV matches very well with the range of excitations in PbS.[15] It is worth highlighting that the DOS of PbS is dominated by Pb vibrations below ~12 meV and by S vibrations above ~12 meV.[15] We conjecture that this spectral separation is the reason for the renormalization of the OA/$d_{14}$-hexane modes towards lower energies, as the OA molecules are attached to the predominantly Pb terminated NCs.[47]

These properties are highlighted in **Figures 6c-d**, which report the difference GDOS (ΔGDOS) of PbS/OA/$d_{14}$-hexane and OA/$d_{14}$-hexane at different temperatures. In the temperature-regime of solid $d_{14}$-hexane, the GDOS of PbS/OA/$d_{14}$-hexane outnumbers the modes in OA/$d_{14}$-hexane below 18 meV and *vice versa* above 18 meV. Upon heating, the mode differences become progressively less pronounced. Above the melting point of $d_{14}$-hexane (178 K), the clearly textured profile of the ΔGDOS changes into a monotone smooth distribution of weak intensity variation at the highest *T*. We corroborate this finding further in **Figure S3** of the Supporting Information by integrating ΔGDOS over three specific energy ranges and find that at highest *T* the integrated signals approach zero. The loss of ΔGDOS contrast in the liquid state signifies the progressive decoupling of the OA/$d_{14}$-hexane dynamics from the NCs. The loss of texture highlights the reduction in the lifetime of vibrational excitations and promotes thus a closer matchable spectral distributions of the solvent with and without PbS NCs.

**Discussion.** This work provides a comprehensive analysis of the structure, dynamics and vibrational properties of the OA ligand shell of PbS NCs in solution. The SANS and SAXS data (**Figure 1**) illustrate that the structure of the OA-ligand sphere resembles the previously described "spiky ball" or "coiled spiky ball" model. This picture is different from the solid state where the collapsed "hair ball"-morphology is believed to be the more stable configuration and a ligand shell thickness of approx. 1.0 nm was measured.[17] The QENS data (**Figure 2**) reveal that the OA molecules exhibit a dynamic adsorption/desorption equilibrium at the surface of the NCs, which manifests itself in a slow, collective center-of-mass diffusion with the NCs and a fast internal diffusion of desorbed, individual OA molecules (**Figure 3**). Both processes can be described with simple Brownian motion. The center-of-mass diffusion and its temperature-dependence are only 50 % smaller than predicted by the Stokes-Einstein model and in good agreement with previous $^1$H-DOSY experiments.[34] These results suggest that the orders of magnitude smaller diffusion constants obtained for NCs in electron microscopy studies originated from the confined sample volume (~100 nm thickness) and was thus not intrinsic to the material.[25–30] The remaining deviation from the expected dilute-limit Stokes-Einstein diffusion may be due to crowding effects.[33] We find evidence for such crowding in the fit of the SAXS/SANS data in **Figure 1**, which indicates the co-existence of individual NCs and clusters of NCs formed by diffusion-limited cluster aggregation.

The internal diffusion of OA in the presence of the NCs is of similar magnitude to that of free, pure OA, however, it is spatially confined to a radius of <10 Å (**Figure 4**), indicating a strong interaction with the NC surface. The fixed window scans (**Figure 5a+b**) further confirm such interaction, which manifests itself in a complete suppression of the phase transformation at ~230K otherwise observed for pure OA in $d_{14}$-hexane. This indicates that the formation of OA clusters to nucleate crystallization is indeed inhibited by the NCs. The very small center-of-mass diffusion constant found only for pure OA in $d_{14}$-hexane (**Figure 3**) is a sign for vesicle formation.

The vibrational properties in **Figure 6** show clear signs of hybrid modes formed by the surface of the PbS NCs and OA. From their mean energy, it follows that these hybrid modes have mainly Pb-character, which is consistent with previous simulations of preferred OA-binding to Pb surface sites.[48] The nanocrystalline character of PbS becomes apparent in the ΔGDOS maximum at energies as low as ~3 meV (**Figure 6d**), which are the transverse acoustic modes of Pb at the surface of the NCs.[15] These modes appear at slightly lower energies than those reported for 1,2-ethanedithiol functionalization of PbS NCs, indicating an even softer NC surface. Due to the smaller spring constant, this should lead to strong electron-phonon coupling, including the mostly unwanted by-pass of a phonon bottleneck.[13] To prevent this coupling, ligands which mechanically stabilize the surface of the NCs would be beneficial, which is an important conclusion of this study.

**Conclusion.** We have studied the temperature-dependent diffusion and phonon spectrum of PbS nanocrystals surface-functionalized with oleic acid in $d_{14}$-hexane solution by neutron backscattering, time-of-flight spectroscopy and small-angle scattering. Our results indicate Brownian diffusion of the nanocrystals and a structure of the ligand shell which differs from that in the solid state. Desorbed ligand molecules also follow simple Brownian diffusion, but their motion is confined to a mean radius < 10 Å, and freezing is inhibited due to the presence of the nanocrystals. The phonon spectrum of the ligand shell exhibits hybrid modes, indicating a preferential binding to Pb-surface sites. Strong low energy phonon modes are consistent with large mechanical softness of the NC surface, which in comparison to recent studies in the solid state appears to be even more pronounced in solution.

**Supporting Information**. The following files are available free of charge: (1) Methods, (2) Details on Data Fitting, (3) Calculation of the scattering power, (4) Calculation of the concentration of free oleic acid in solution, (5) Estimation of the number of desorbed ligands, (6) Calculation of the volume fraction, (7) Temperature-dependent redistribution of phonon modes, (8) Characteristic vibrational energies in the solid and liquid state, Figure (S1): Qualitative scheme describing the results of the SAXS/SANS fitting, Figure (S2): Scattering power values, Figure (S3): Temperature-dependent relative difference in the number of phonon modes, Figure (S4): GDOS of $d_{14}$-hexane at 100 K, Figure (S5+S6): QENS data from IN5.

**Corresponding Author:** *marcus.scheele@uni-tuebingen.de

**Funding Sources.** We are grateful for financial support by the DFG under grants SCHE1905/4 and SCHR700/25. This work has received funding by the European Research Council (ERC) under the European Union's Horizon 2020 research and innovation program (grant agreement No 802822).

Notes. The neutron data can be accessed at: http://doi.ill.fr/10.5291/ILL-DATA.9-10-1514 (IN5) and http://doi.ill.fr/10.5291/ILL-DATA.9-12-498 (IN16B and D11).

**Acknowledgements.** We wish to thank the ILL for making their excellent facilities available to us. FS wishes to thank the ILL for their hospitality during his sabbatical stay. The ID02 team at ESRF, especially Theyencheri Narayanan, is gratefully acknowledged for providing in-house beam time.

# Supporting Information

**Table of Content**

# Methods

*Synthesis of PbS/OA NCs*

PbS NCs are synthesized by a procedure modified from the work of Weidman *et al.*[1] 0.040 g of sulfur and 7.5 mL of OLA are filled into a 20 mL glass vial and placed on a magnetic stirrer at room temperature in a nitrogen filled glovebox overnight. 7.5 g of $PbCl_2$ and 22.5 mL of oleylamine are filled into a 100 mL three neck flask equipped with a reflux condenser, a septum and a temperature control couple. The flask is evacuated for ~ 30 minutes until the bubble formation in the liquid stops, after which the mixture is heated to 120 °C under nitrogen. Care must be taken in this step that the $PbCl_2$ powder stays well suspended in the oleylamine. 6.75 mL of the sulfur-oleylamine solution are swiftly injected. The temperature subsequently drops to ~ 100 °C, but rises to 120°C within 1-2 min. An overshoot of temperature is prevented by tuning down the thermostat shortly before reaching 120 °C. After the reaction time (typically 30 min), the mixture is rapidly cooled down by replacing the heating mantle with a water bath and injecting 60 mL of hexane. The reaction mixture is transferred to a nitrogen-filled glovebox and precipitated with ~ 80 mL ethanol. The suspension is centrifuged with 4000 rpm for 5 minutes and the supernatant is discarded. The precipitate is redispersed in ~ 80 mL hexane. There is still unreacted $PbCl_2$ in that mixture that needs to be separated from the QD solution. That is done by centrifuging it again at 4000 rpm for 5 minutes and discarding the white precipitate. The NCs are again precipitated by ~ 80 mL of ethanol and after discarding the supernatant, 4 mL of degassed oleic acid (OA, Aldrich, 90 % technical grade) are added and the mixture is stirred with a spatula. After leaving the mixture stand for minimum 1 h, it is cleaned three times by a.) adding ~ 80 mL hexane, b.) adding ~ 80 mL of ethanol, c.) centrifuging at 4000 rpm for 5 minutes and discarding the supernatant after centrifugation.

*SANS and SAXS*

SAXS experiments were performed on the instrument ID02 at ESRF - The European Synchrotron, Grenoble, France.[2] A CCD detector Rayonix MX170HS was used with binning 2×2 (effective square pixel size 88.34 µm corresponding also to the point-spread-function characteristic width), placed at 3 distances from the sample: 0.77, 5.00 and 30.69 m (sample offset measured with silver behenate). The wavelength was constant at 0.0990 nm (12519.0 eV as regularly verified from absorption edges of metal foils, relative fwhm $10^{-4}$). The illuminated sample cross-section was about 0.4x0.6 mm² at high q and 0.25x0.4 mm² at mid and low q, with a Gaussian profile and most of the flux density on 50x150 µm². The flux was $5.5*10^{12}$ ph/s at high q and $2*10^{12}$ ph/s at mid and low q, with exposure times ranging from 30 to 250 ms. Samples were poured in quartz capillaries (WJM-Glas, Berlin, Germany) of ca. 1.5 mm pathway and 10 µm wall thickness. Data were automatically corrected with the beamline's standard workflow accounting for transmitted photons measured with a calibrated PIN diode atop the beam stop (with a known delay between fast shutter and detector acquisition), flat field, dark, spatial distortion; 2D data were azimuthally averaged. About 10 frames were averaged and the standard-deviation was used as error-bar. Intensities were corrected by the capillary thickness determined by transmission scan. The contribution from the solvent was subtracted. The absolute scale was cross-checked by comparison of the forward scattering of the solvent and the theoretical value from isothermal compressibility and electron density and was better than 1 %.

SANS experiments were performed on the small-angle scattering instrument D11 (Institut Laue-Langevin (ILL), Grenoble, France).[3] Samples were filled into quartz cells (110-QS, Hellma, Müllheim, Germany) and the cells were placed on a temperature-controlled copper sample holder. The measurements were performed at room temperature.

The samples were measured at sample-to-detector distances of 39, 8 and 1.4 m, covering a $q$ range from 0.0015 to 0.45 Å$^{-1}$. The incoming neutron wavelength, $\lambda$, was 6 Å with a full width-half maximum (FWHM) wavelength spread of 9%. The beam size was 7 x 10 mm². Scattered neutrons were detected via a $^3$He gas detector (CERCA) with a pixel size of 3.75 x 3.75 mm² and a total pixel number of 256 x 256. Data were calibrated to an absolute scale using water (H$_2$O) scattering intensity, $d\sigma/d\Omega = 0.983$ cm$^{-1}$ as a secondary calibration standard. Raw data were saved in the NeXus (.nxs) format.[4] Prior to further analysis, all 2D scattering profiles obtained were corrected for both transmission and background scattering

### *Quasi-Elastic Neutron Scattering (QENS)*

QENS spectra were recorded on the IN16B and the IN5 spectrometers at the ILL and saved in the NeXus (.nxs) format.[4] The maximum energy transfer (30 µeV) and energy resolution (0.8 µeV) at IN16B are ideally suited to study diffusive processes in the range for $D = 10^{-10} - 10^{-12}$ m²/s.[5] Full spectra (energy range -30 µeV < E < +30 µeV) were measured on IN16B at the temperatures 183K, 239K, and 295K. In addition, elastic ($\Delta E = 0$ µeV) and inelastic ($\Delta E = 1.3$ µeV) fixed window scans were recorded in the temperature range 239K<T<295K.[6] This experiment achieves a very high energy resolution by defining the incident neutron energy and determining the scattered neutron energy using Bragg scattering at perpendicular incident angles to the monochromator and analyzer single crystal surfaces. The backscattering single crystals were chosen to be Si(111), corresponding to E = 2.08 µeV, employing the cold neutron backscattering spectrometer IN16B at the ILL, achieving an energy resolution of 0.8µeV FWHM.[5]

In contrast, the energy resolution at IN5 is sufficient to resolve the QENS signal of fast relaxational dynamics, such as diffusion processes with diffusion coefficients $D$ significantly larger than 10$^{-10}$ m²/s. This includes the diffusion of the d$_{14}$-hexane solvent. To this end, we used the Stokes and anti-Stokes lines of the INS signal acquired at IN5 as laid out below.

### *Inelastic Neutron Scattering (INS) by Time-of-Flight Spectroscopy (TOF)*

The neutron energy loss (Stokes line) is limited to less than the incident neutron energy, which was chosen to be 3.3 meV ($\lambda = 5$ Å) at the TOF spectrometer IN5 at ILL. With this setup, we accessed an energy resolution of 80 µeV FWHM and a momentum $q$ range of 0.3-2.5 Å$^{-1}$ at the elastic line. There is no limit to the energy gain of neutrons (anti-Stokes line), however, the signal intensity is scaled by the Bose occupation number. Thus, the signal quality is temperature-dependent.[5] All samples were measured at 100, 150, 183, 200, and 239 K. The temperatures employed in our experiment were high enough to record the inelastic response in the entire energy range of vibrational excitations of the PbS NCs, i.e. more than 35 meV. In addition, PbS/OA/$d_{14}$-hexane and OA/$d_{14}$-hexane were measured at 2 K for low-temperature resolution and background references. A standard helium cryostat was used with He exchange gas of a pressure of about 10 mbar at 100 K for thermalization purposes.

*Data Reduction and Analysis*

At IN5 and IN16B, auxiliary correction measurements of empty sample holder and vanadium standard have been carried out with the sample scans. Standard data correction for background scattering, detector efficiencies comprising the energy dependence of the counter efficiencies were applied to the IN5 data. The IN5 signal recorded as a double differential cross section in the natural units of the experiment, i.e. TOF and scattering angle 2θ, was sequentially transformed into the dynamic structure factor $S(2\theta,\omega)$, with ω denoting energy, then into the GDOS(2θ,ω) and summed over spectra in the 2θ range of 50 to 130 degrees. The derived GDOS has been normalized to 60 phonon modes in the energy range 0.5-48 meV irrespective of the sample composition. This approach is justified by the negligible scattering contribution of the PbS NC in the specimen. The data were reduced with the software package lamp and the standard routines implemented therein (https://www.ill.eu/users/support-labs-infrastructure/software-scientific-tools/lamp/).

For the analysis of the IN5 QENS signal, standard reductions and the interpolation of $S(2\theta,\omega)$ onto a constant momentum *q* grid as $S(q,\omega)$ were carried out using the software package lamp (https://www.ill.eu/users/support-labs-infrastructure/software-scientific-tools/lamp/).

Equivalent standard reductions were carried out for the IN16B QENS data, using the software package Mantid (www.mantidproject.org). No interpolation to a constant (*q*,ω)-grid was required in this case due to the small energy transfers.

The QENS data from both IN16B and IN5 were subsequently fitted using the python3 scipy.optimize.curve_fit algorithm, employing the models as reported in the results section and SI.

# Fitting of the QENS Data from IN16B

The OA accounts for most of the incoherent scattering reflected by the QENS signals in **Figure 2** in the main body of the manuscript. Due to the low OA concentration, these measured signals, even though well visible, are too weak to be reliably fitted for each *q*-value independently. For this reason, we include the *q*-dependence of the scattering function in our model and fit the spectra for all (*q*,ω) simultaneously according to the following heuristic model for the measured intensity $S(q,\omega)$:[7,8]

$$S(q,\omega) = R \otimes \{\beta(q)[A_0(q)L(\gamma(q),\omega) + (1 - A_0(q))L(\gamma(q) + \Gamma(q),\omega)] \\ + \beta_{solvent}(q)L(\gamma_{solvent}(q),\omega) + \beta_\delta(q)\delta(\omega)\} + a\omega + b \quad (1),$$

where $R = R(q,\omega)$ is the spectrometer resolution function, $L(\cdot,\sigma)$ a Lorentzian function with the half-width at half-maximum $\sigma$, $\delta(\omega)$ the Dirac delta-function describing an apparent elastic contribution, and $0 \leq A_0(q) \leq 1$, $\gamma(q) \geq 0$, $\Gamma(q) \geq 0$, $\beta(q) \geq 0$, $\beta_\delta(\omega) \geq 0$, and $\beta_{solvent}(q) \geq 0$ are scalar fit parameters.

This model accounts for a superposition of a center-of-mass diffusion of the PbS NCs (or alternatively OA vesicles) and a superimposed - i.e. convoluted - internal diffusion of OA decorating the NCs. The center-of-mass diffusion is associated with a line broadening of the QENS signal by width $\gamma(q)$, while the molecular diffusion is associated with the line width $\Gamma(q)$. The convolution of the two processes is implemented by the summation $\gamma(q) + \Gamma(q)$. The much faster $d_{14}$-hexane solvent molecule diffusion is described by the third Lorentzian contribution with the width $\gamma_{solvent}(q)$. In addition, a possible contribution from very large objects that appear immobile on the observation time scale of IN16B is accounted for by the last term containing the Dirac delta-function. This last term simultaneously accounts for any imperfect subtraction of the container signal. At the lowest $q$, this term may also account for the small-angle scattering from the decorated nanoparticles. Finally, the fitted model allows for an apparent sloped background, $a\omega + b$.

We fit both the PbS/OA/$d_{14}$-hexane and the OA/$d_{14}$-hexane IN16B spectra by equation (1). In these fits, the parameters $\beta_{solvent}(q)$, $\gamma_{solvent}(q)$, $a$, and $b$ are fixed using the results of the corresponding pure solvent $d_{14}$-hexane fits, with $\beta_{solvent}(q)$ being rescaled accounting for the volume excluded by the PbS. $A_0(q)$ can be identified with the Elastic Incoherent Structure Factor (EISF) associated with the oleic acid diffusion. Since $\beta(q)$ and $\beta_{solvent}(q)$ are free fit parameters, the fit results for the linewidths $\gamma(q)$, $\Gamma(q)$, and $\gamma_{solvent}(q)$, as well as for $A_0(q)$ are not sensitive to the exact normalization of the apparent detector efficiency. Technically, the convolution $R \otimes \cdot$ is carried out by modeling R as a sum of an arbitrary number of Gaussian functions (3 in the case of IN16B, 5 for IN5). Therefore, the observable $S(q,\omega)$ can be fitted by a sum of Voigt functions.

The contribution of the $d_{14}$-hexane solvent manifests itself as a broad apparent background due to its fast diffusion. The independent fits to this solvent signal, resulting in $\beta_{solvent}(q)$ and $\gamma_{solvent}(q)$, are represented by dash-dotted lines. Note that the plots show binned data for better visibility, but the fits have been performed prior to binning for higher accuracy.

## Fitting of the QENS Data from IN5

The model function for the quasi-elastic scattering modeled on IN5 consists of a sum of two Lorentzian functions $L_1$ and $L_2$ accounting for the solvent motions and a Dirac $\delta$-function accounting for the apparent elastic scattering. This model is inspired by a model for pure water:[9]

$$S(q,\omega) = R(q,\omega) \otimes [I_1(q)L_1(\sigma_1(q),\omega) + I_2(q)L_2(q,\omega) + I_3(q)\delta(\omega)] \quad (2)$$

Therein, the intensities $I_{1,2,3}$ and the Lorentzian widths $\sigma_{1,2}(q)$ are scalar fit parameters, $R$ the spectrometer resolution function, and the $\otimes$-symbol denotes the convolution. As described for the IN16B QENS data, the convolution of the IN5 QENS model function with the resolution function was carried out analytically by employing Voigt functions. The EISF and nanosecond QENS signals seen on IN16B completely "disappear" within the broad resolution on IN5. Therefore, these contributions observed on IN16B are represented by the apparent elastic contribution ("Dirac" $\delta$-function in equation (2)) in the model scattering function for IN5. Conversely, the broad signal from the fast solvent contribution seen on IN5 appears as an apparent nearly flat background or very broad Lorentzian on IN16B at the limit of the accessible energy range of IN16B of 30µeV.

## Analysis of SANS and SAXS Data

### A Brief Introduction into Small-Angle Scattering

Generally, the angle-dependent scattered intensity, *I(q)*, obtained from a small-angle scattering experiment is approximated by the following equation, which is correct for monodisperse centro-symetrically interacting scatterers:[10]

$$I(q) \sim n \cdot P(q) \cdot S(q)$$

where $q = |\vec{q}| = \frac{4\pi}{\lambda}\sin\theta$ (with the incoming wavelength $\lambda$ and the scattering angle $2\theta$) is the momentum transfer, $n$ is the particle number density, *ΔSLD* is the contrast (scattering length density difference) between solvent and solute and $V$ is the volume of one solute particle. The term *P(q)* is the unnormalized ($P(0) = (\Delta SLD \cdot V)^2$) form factor and represents the Fourier transform of the scattering length density of an individual particle, describing its overall shape and internal heterogeneities.[11] The term *S(q)* is the structure factor and corresponds to the Fourier transform of the pair correlation function of the system in question; it describes the net overall interactions between solute particles interacting with each other.

### Model-free information obtained from SANS and SAXS data

In the case of spherical particles, their radius $R$ can be determined in a model-free fashion from the minima of the *I(q)* oscillations, $q_{min}$, *via* the relation

$$\tan(q_{min}R) = q_{min}R$$

i.e., $q_{min} \cdot R = 4.49, 7.73,...$[12] The radii of the PbS NCs are discussed in the main text.

## SANS and SAXS Data Fitting

SANS and SAXS data were fitted with the "SASfit" software package.[13] Unless specified otherwise, all following equations are taken from the SASfit manual written by J. Kohlbrecher. The values of all fit parameters for both the SAXS and SANS data sets are listed in **Table S1**.

The **background scattering** was accounted for by the constant ($q$-independent) contribution $I_{bkg}$. It contains both the incoherent scattering due to mostly hydrogen for SANS data, and the solvent compressibility term for SANS and SAXS.

The **core-shell spherical form factor, $P_{cs}(q)$**, was used to describe oleic acid coated PbS nanoparticles. It is defined as follows:

$$P(q, R, t, SLD_c, SLD_s, SLD_0) = [K(q, R + t, SLD_s - SLD_0) + K(q, R, SLD_c - SLD_s)]^2$$

with

$$K(q, R, \Delta SLD) = \frac{4}{3}\pi R^3 \Delta SLD \cdot 3\frac{\sin qR - qR \cos qR}{(qR)^3}$$

Here, $R$, $t$, $SLD_c$, $SLD_s$, $SLD_0$ are the radius of the core, the shell thickness, the scattering length density of the core (index "c"), shell (index "s") and solvent (index "0").

The form factor was complemented by an ad hoc Ornstein-Zernike (OZ) contribution:

$$I_{OZ}(q) = \frac{I_{OZ}(0)}{1 + q^2\xi_{OZ}^2}$$

where $\xi_{OZ}$ is the correlation length. This additional contribution was found necessary for SANS and SAXS data, with the same correlation length for both techniques and an intensity proportional to the contribution from all nanoparticles, and seems therefore linked to some features of these nanoparticles not accounted for by the core-shell spherical model.

A spherical **mass fractal structure factor, $S(q)$,** with exponential cut-off was used, described by the following formula:

$$S(q) = 1 + \frac{D_f}{r_0^{D_f}} \int_0^\infty r^{D_f - 3} h_{Exp}(r, \xi_f) \frac{\sin(qr)}{qr} r^2 dr,$$

The exponential cut-off function, $h_{Exp}(r,\xi)$ is defined *via*

$$h_{Exp}(r, \xi_f) = \exp\left[-\left(\frac{r}{\xi_f}\right)^\alpha\right].$$

$r_0$, $\xi_f$, $D_f$ and $\alpha$ are the characteristic dimension of the individual scattering objects, the cut-off characteristic length for the fractal correlations, the fractal dimension and the exponent of the exponential cut-off function, respectively. Their respective values are given in **Table S1**. As a reminder, for the case of diffusion-limited cluster aggregates (DLCA) the fractal dimension is around 1.8 while for reaction-limited cluster aggregates (RLCA) the fractal dimension is around 2.1, and the exponent $\alpha$ is empirically found to be close to 2 in either case.[14]

Within a $q$ range insensitive to the form factor and the internal structure of the system under study, the dependence of the small-angle scattering intensity on $q$ can be expressed[15] *via* the power law

$$I(q) \sim q^{-D_f}$$

where the exponent $D_f$, in the case of a fractal (i.e., self-similar) system, corresponds to the so-called fractal dimension quantifying the manner in which the mass of the fractal structure increases in space.[16] $D$ values with $3 \leq D \leq 4$ describe so-called "surface fractals", whereas $D < 3$ describes mass fractals.[17]

The SAXS data in **Figure 1** feature a power-law decay close to $q^{-1}$, which seemingly reflects the presence of rod-like (linear) aggregates ($D_f \approx 1$). The mass fractal fit to the SANS data, however, yields $D_f = 1.85$, i.e. SANS and SAXS (with their different contrasts towards PbS and oleic acid) apparently disagree. Given the stability of the samples, we deduce that the scattering data actually emerge from the sum of two additive and independent contributions: that of free (dispersed, non-interacting) PbS nanoparticles plus a contribution from aggregated PbS nanoparticles. The different intensity contributions for SANS and for SAXS can then be explained by a different PbS/oleic acid ratio for free particles and for aggregated particles, with fractal aggregates depleted in oleic acid, and therefore less visible by SANS.

The model finally used is the following:

$$I(q) = I_{bkg} + I_{OZ}(q) + I_{cs,free}(q) + I_{cs,agg}(q)$$

where $I_{cs,free}(q)$ describes the contribution from non-interacting particles (no structure factor) and $I_{cs,agg}(q)$ describes the contribution of particles in fractal aggregates. The same form factor is used for both free and aggregated nanoparticles, except for the solvation of the shell that is let free. The same parameters are used to fit the SANS and SAXS data, except for the scattering length densities that are fixed to their respective known values (the shell SLD is recalculated from oleic acid and solvent SLDs based on solvation).
Qualitatively, the model describes a system sketched in **Figure S1**.

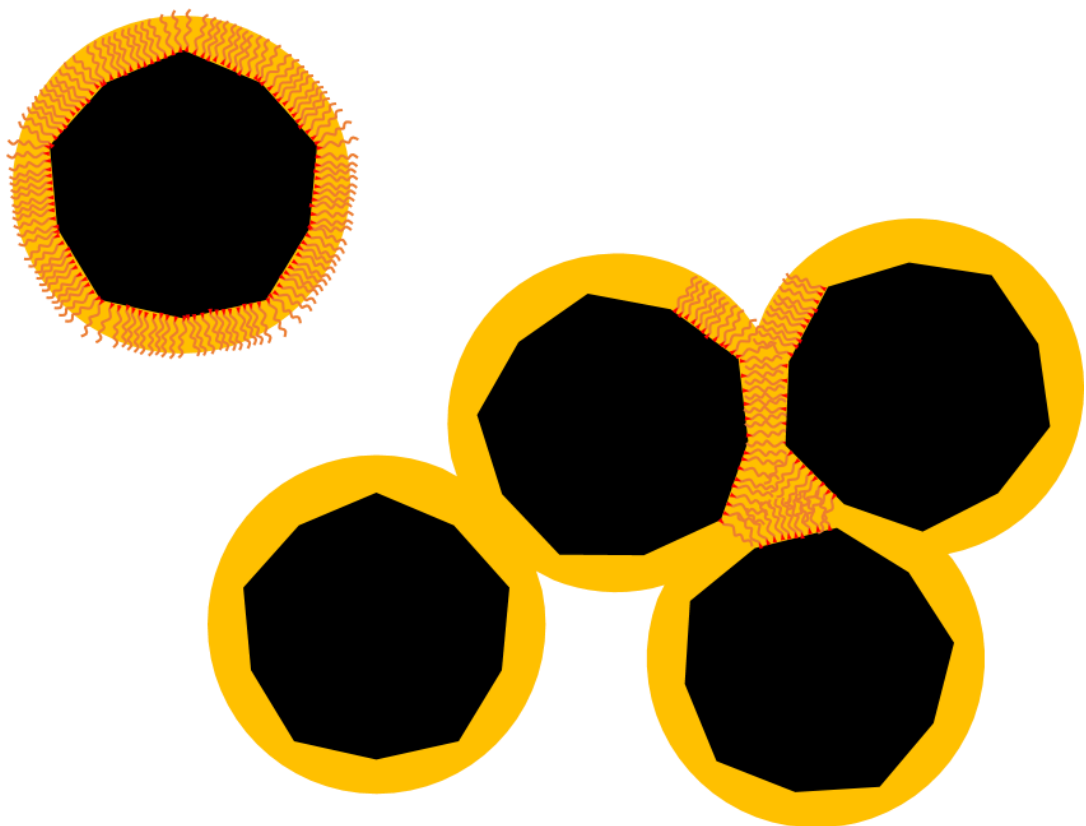

**Figure S1**: Illustration of the system modeled for the analysis of the SAXS data. Quasi-spherical PbS nanoparticles (black) are surrounded by an oleic-acid shell. Both free and aggregated nanoparticles are present simultaneously. The aggregated nanoparticles are depleted in oleic acid compared to the free nanoparticles, which may be justified by the sharing of oleic acid shells between neighboring particles in the fractal aggregate.

For the monomers, we calculate the degree of solvation of the OA shell to be ~80% (details of this calculation are given below). For the aggregates, we observe an OA/PbS ratio which is ~33% smaller than the one for the monomers, suggesting that the OA shells of the individual monomers overlap inside the aggregates. Whether the depletion in oleic acid for aggregated particles is a cause or a result of aggregation is unknown.

As an additional piece of information, the $D_f$ value of 1.85 obtained from SANS indicates diffusion-limited cluster aggregation (DLCA) to be the preferential mechanism of aggregate formation.[18] This is consistent with the deviation from the expected dilute-limit Stokes-Einstein diffusion observed *via* QENS.

**Table S2**. *Parameters obtained from SANS and SAXS fits using sasfit. (\* indicates mean radius, with a log-normal distribution of relative standard-deviation 5.5 %)*

| Contribution | Parameter | SANS | SAXS | Unit |
|---|---|---|---|---|
| **Background** | $I_{bkg}$ | 0.047 | 0.024 | cm$^{-1}$ |
| **Form factor** | $R$ | 33* | | Å |
| | $t$ | 18 | | Å |
| | $\Delta SLD_c$ | 3.8·10$^{-6}$ | -41.4·10$^{-6}$ | Å$^{-2}$ |
| | $\Delta SLD_s$ (solvation) | | | Å$^{-2}$ |
| | Dispersed NP | 4.8·10$^{-6}$ (80 %) | -1.0·10$^{-6}$ (80 %) | |
| | Aggregated NP | 3.0·10$^{-6}$ (50 %) | -0.5·10$^{-6}$ (50 %) | |
| **Ornstein-Zernike** | $I_{OZ}(0)$ | 2 | 5 | cm$^{-1}$ |
| | $\xi_{OZ}$ | 36 | | Å |
| **Structure factor** | $r_0$ | 35 | | Å |
| | $\xi_f$ | 310 | | Å |
| | $D_f$ | 1.85 | | |
| | $\alpha$ | 1 | | |
| **Fraction aggregated** | $n_{cs,agg}/(n_{cs,agg}+n_{cs,free})$ | 21 % | | |

Calculation of Solvation Degree of OA shell

The conclusion that 80 % of the OA molecules in the shell are solvated is obtained as follows:

The theoretical neutron scattering length density difference, between d14-hexane and oleic acid is (for the respective values please see **Table S2** below)

$$SLD_{dhex} - SLD_{OA} = (6.10 \cdot 10^{-4} - 0.08 \cdot 10^{-4}) \frac{1}{\text{Å}^2} = 6.02 \cdot 10^{-4} \frac{1}{\text{Å}^2}$$

The fit to the SANS data for dispersed nanoparticles yields $SLD_s = 4.81 \cdot 10^{-4} \frac{1}{\text{Å}^2}$, which corresponds to ~80% of the theoretical value calculated above, thereby indicating a 20 % solvation degree of the OA shell (the SLD of the shell for SAXS also accounts for this solvation). For aggregated nanoparticles, the same calculation leads to a solvation of 50 %, i.e. the surface coverage by ligand is less for aggregated particles.

## Scattering Length Densities

The scattering length densities (SLD) of the components of the system were calculated using the online tool *sld-calculator.appspot.com*. The respective values are given in **Table S2**.

**Table S3**. *Scattering length densities of the respective system components.*

| Component | Density (g/cm³) | X-Ray SLD (1/ Å²) | Neutron SLD (1/ Å²) |
|---|---|---|---|
| Oleic acid ($C_{18}H_{34}O_2$) | 0.895 | $8.50 \cdot 10^{-6}$ | $0.08 \cdot 10^{-6}$ |
| d14-hexane ($C_6D_{14}$) | 0.763 | $6.46 \cdot 10^{-6}$ | $6.10 \cdot 10^{-6}$ |
| PbS | 7.6 | $47.9 \cdot 10^{-6}$ | $2.34 \cdot 10^{-6}$ |

## Calculation of the Scattering Power

*Concentration of H-, D- and C-atoms in pure $d_{14}$-hexane*

With a molar weight of 100.3 g/mol and a density of 0.767 g/mL, the concentration of $d_{14}$-hexane in $d_{14}$-hexane is 7.65 mol/L. With 14 deuterium atoms per $d_{14}$-hexane, the concentration of D-atoms in pure $d_{14}$-hexane is 107.1 mol/L and the concentration of C-atoms is 45.9 mol/L. With an isotopic purity of 1 %, the concentration of H-atoms is 1.07 mol/L.

*Concentration of H-, D-, C- and O-atoms in 35 mmol/L of oleic acid in $d_{14}$-hexane*

The concentration for oleic acid of 35 mmol/L was chosen to match the concentration of bound oleic acid in the PbS/OA/$d_{14}$-hexane solution. With 34 H-atoms, 18 C-atoms and 2 O-atoms, this adds 1.19 mol/L H-atoms, 0.63 mol/L C-atoms, 0.07 mol/L O-atoms and 0 mol/L D-atoms to the results for pure $d_{14}$-hexane. The total concentrations in OA/$d_{14}$-hexane are therefore 107.1, 2.26, 46.5 and 0.07 mol/L for D, H, C and O, respectively.

*Concentration of H-, D-, C-, O-, Pb- and S-atoms in 140 µmol/L of PbS/OA in $d_{14}$-hexane*

There are roughly 4000 atoms in a PbS NC with a radius of 3.3 nm.[19] The excess of Pb over S-atoms (due to Pb ad-atoms in the ligand sphere) is roughly 1.3. Therefore, the concentrations of Pb and S in a 140 µL solution are 0.32 mol/L and 0.24 mol/L, respectively. Assuming a spherical particle size and a coverage of 3 OA molecules per nm² results in the same concentration of OA as in the previous section. Thus, the concentration of the D-, H-, C- and O-atoms is approximately the same. The total concentrations in PbS/OA/ $d_{14}$-hexane are 107.1, 2.26, 46.5, 0.07, 0.32 and 0.24 mol/L for D, H, C, O, Pb and S, respectively.

Multiplying these values with the total scattering cross-sections of the elements (7.6 * $10^{-24}$ cm$^2$, 82 * $10^{-24}$ cm$^2$, 5.6 * $10^{-24}$ cm$^2$, 4.2 * $10^{-24}$ cm$^2$, 11 * $10^{-24}$ cm$^2$ and 1.0 * $10^{-24}$ cm$^2$) yields the scattering power relevant for the time-of-flight data of all elemental components in the NC sample as shown in the **Table S3**.[20] Note that for the backscattering data only the incoherent part of the scattering cross section is to be considered.

*Table S4. Relative scattering power of the elements in the studied samples*

| Element | D | H | C | O | Pb | S |
|---|---|---|---|---|---|---|
| Norm. Scattering Power | 3392 | 772 | 1085 | 1.25 | 14.7 | 1 |

**Figure S2** illustrates these relationships for all elements on a logarithmic scale.

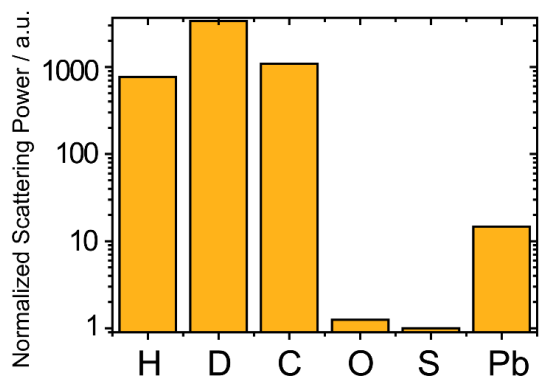

**Figure S2**. Scattering power of all elements in the sample according to the calculation in the text.

# Calculation of the concentration of free oleic acid in the NC sample

Density Functional Theory calculations have shown that the binding strength of OA to the {100}-facets of PbS is relatively weak (160 meV) with an adsorption constant of 500 L/mol.[19,21,22] The density of oleic acid on the surface of the NCs is roughly 3 nm$^{-2}$, which results in a coverage of ~400 OA molecules per NC.[19] Based on the assumption that every second OA molecule resides on a {100}-facet and that OA molecules bound to other facets never desorb, there are 200 weakly bound OA molecules per particle. With an NC concentration of 140 µmol/L, the total concentration of exchangeable OA in the NC sample is 28 mmol/L with an adsorption constant of 500 L/mol. This corresponds to 56 µmol/L desorbed OA molecules in the sample. Thus, there are approximately three NCs per desorbed OA molecule at any time. This explains why we also observe the expected diffusion coefficient of free OA in the NC sample ($D_2$).

# Estimation of the number of ligands to desorb from the NC surface within the measurement time window

Assuming that ligand desorption follows first order kinetics, the desorption rate is

$$v_{des} = k_{des} \cdot [OA_{bound}]$$

We infer an estimate for $k_{des}$ from previous studies on the rate of desorption for amine-stabilized CdSe NCs in solution as $(10^{-2} - 10^{-4})$ s$^{-1}$.[23,24] As shown above, $[OA_{bound}]$ = 28 mmol/L, under the assumption that only OA molecules bound to 100 facets desorb into the liquid phase. Thus, we estimate $v_{des} = (3 \cdot 10^{-4} - 3 \cdot 10^{-6}) mol L^{-1} s^{-1}$.

Considering that the sample volume investigated during a QENS experiment is roughly 1 mL and the integration time per measurement is on the order of 1 ns, this allows us to gauche the total number of desorbed OA molecules in the sample during a single measurement as

$$N = N_A \cdot (3 \cdot 10^{-4} - 3 \cdot 10^{-6}) mol L^{-1} s^{-1} \cdot 10^{-3} L \cdot 10^{-9} s = 2 \cdot 10^8 - 2 \cdot 10^6$$

# Calculation of the volume fraction in the NC sample

The total volume of a PbS NC with 3.3 nm radius and the density of PbS (7.6 g/cm$^3$) is 150 nm$^3$. The volume of the ligand sphere can be approximated from its width of 1.8 nm (**Figure 1**) and a van-der-Waals radius of the backbone of roughly 2.5 Å, which gives 0.1 nm$^3$. With ~400 OA molecules per NC (see previous section), the volume occupied by the ligand sphere is 40 nm$^3$ and, thus, the total volume occupied by the NC and its ligand sphere is 190 nm$^3$. One liter of the NC solution contains 140 µmol of particles, which occupy 16 mL. The volume fraction is therefore 1.6 %.

# Temperature-dependent redistribution of phonon modes in the NC sample

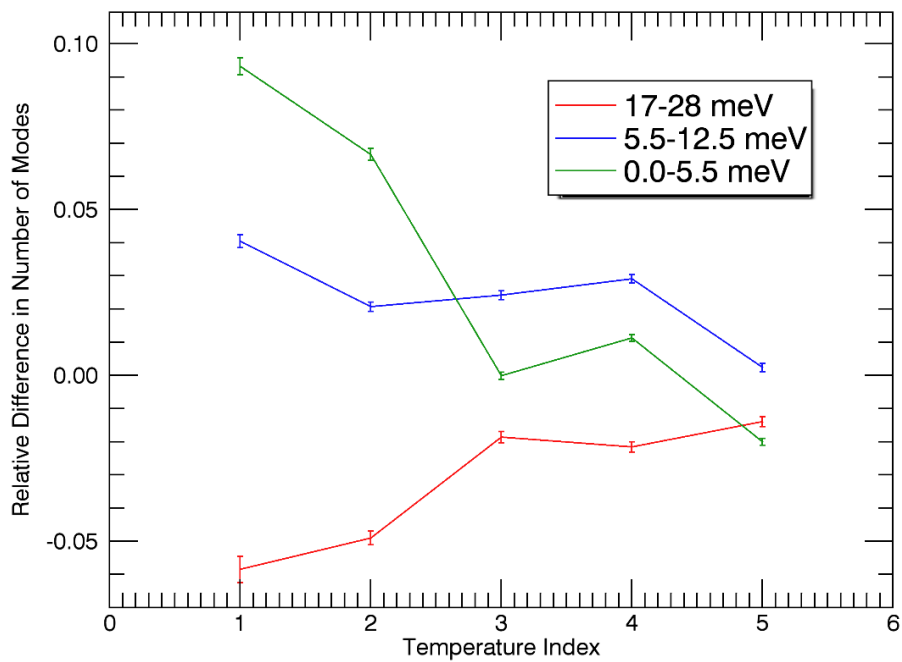

**Figure S3.** Temperature dependence of the relative difference in the number of phonon modes for PbS/OA/$d_{14}$-hexane. The temperature index on the x-axis is: (1) 100 K, (2) 150 K, (3) 183 K, (4) 200 K and (5) 239 K. Phonon modes are binned into the three energy regimes indicated in green, blue and red color. See also Figures 6c-d in the main part for further details.

The redistribution of modes upon heating is indicated by the relative difference in the number of modes (RDNM, **Figure S3**). The RDNM has been calculated for three energy intervals with the most significant features observed in the low and high energy ranges (0-5.5meV, 5.5-12.5 meV and 17-28 meV; see **Figure 6c-d** in the main part). At 100 K, the excess of modes is as high as 10% and 6% in the regime with lowest and highest energy, respectively. Upon heating, the RDNM approach zero, indicating the decoupling of the OA/$d_{14}$-hexane dynamics from the NCs and hence closer matchable spectral distributions of the solvent with and without NCs. The intermediate energy range displays a less characteristic behavior as both positive and negative intensity contributions merge with increasing T.

# Characteristic energies in the solid and liquid states

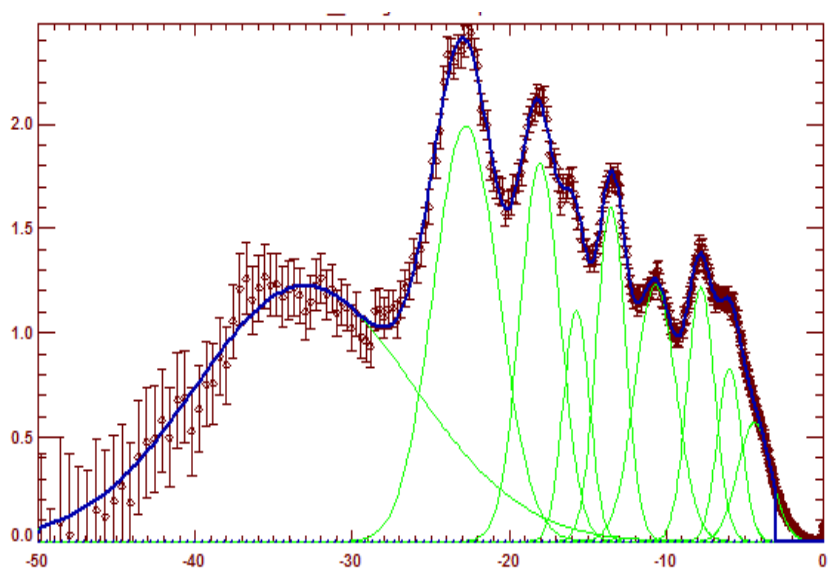

**Figure S4**. GDOS of $d_{14}$-hexane at 100 K. Solid lines correspond to Gaussian fits to the characteristic mode peaks.

The addition of OA and nano-PbS/OA to $d_{14}$-hexane modifies the peak intensities, enhancing particularly the modes in the low-energy range. However, the energies of the characteristic peaks are not altered. We conclude that the mixed specimens crystallize in a structure reminiscent of crystalline $d_{14}$-hexane. To quantify these observations, we approximate the densities of states with a set of Gaussians and list the characteristic energies derived at 100 K and 183 K in **Table S4** and **Table S5**, respectively. The fit quality is indicated in **Figure S4** reporting the set of Gaussians fitted to the GDOS of $d_{14}$-hexane. We find a very good match of the fitted energies with literature data exploiting other spectroscopic techniques.[25,26] Thus, we may safely state that at energies lower than 12 meV (~100 cm$^{-1}$), the mode peaks are exclusively due to collective lattice modes of the nanocrystals. At higher energies, a discrimination between collective lattice and intra-molecular vibrations is made difficult as both signals overlap. The approximation of the few characteristic features in the GDOS of the liquid samples results in energy parameters in agreement with literature data for intra-molecular vibrations. The enhanced variance of the results between the different specimens is the result of the reduced distinctiveness of the features. Its significance has to be taken with care.

**TABLE S4:** *Characteristic energies of the three specimens derived from Gaussian fits to the GDOS data at 100 K, i.e. in the frozen state of $d_{14}$-hexane. The energy and their reliability parameters are reported in $cm^{-1}$ for convenient comparison with literature data.*

| PbS/OA/dH [cm$^{-1}$] | OA/dH [cm$^{-1}$] | dH [cm$^{-1}$] |
|---|---|---|
| 32.8 ± 1.4 | 34.0 ± 0.2 | 33.9 ± 0.4 |
| 49.1 ± 0.3 | 48.6 ± 0.1 | 48.2 ± 0.3 |
| 63.6 ± 0.3 | 63.6 ± 0.1 | 63.2 ± 0.2 |
| 85.2 ± 0.2 | 86.0 ± 0.1 | 86.1 ± 0.3 |
| 109.4 ± 0.2 | 108.6 ± 0.1 | 109.1 ± 0.3 |
| 128.2 ± 0.3 | 127.6 ± 0.2 | 126.7 ± 0.4 |
| 147.8 ± 0.4 | 147.9 ± 0.2 | 145.6 ± 0.4 |
| 183.2 ± 0.2 | 184.8 ± 0.1 | 183.6 ± 0.4 |
| 264.0 ± 1.0 | 262.9 ± 0.6 | 266.3 ± 2.4 |
| 293.1 ± 3.2 | 296.9 ± 1.6 | |
| 338.5 ± 2.1 | 333.7 ± 2.9 | |

**TABLE S5:** *Characteristic energies of the three specimens derived from Gaussian fits to the GDOS data at 183 K, i.e. above the liquidus temperature of $d_{14}$-hexane. The energy and their reliability parameters are reported in $cm^{-1}$ for convenient comparison with literature data.*

| PbS/OA/dH [$cm^{-1}$] | OA/dH [$cm^{-1}$] | dH [$cm^{-1}$] |
|---|---|---|
| 135.3 ± 0.2 | 135.3 ± 0.8 | 140.7 ± 1.1 |
| 165.2 ± 1.4 | 160.9 ± 5.2 | 165.3 ± 1.2 |
| 176.5 ± 0.2 | 182.7 ± 0.9 | 184.9 ± 1.3 |
| 224.4 ± 0.5 | 238.4 ± 1.1 | 235.6 ± 2.6 |
| 290.7 ± 0.4 | 287.0 ± 1.1 | 291.8 ± 1.3 |

## Quasi-elastic scattering recorded on IN5

Due to the large sample container diameter (22mm), the container deteriorates the resolution (**Figure S5a**). **Figure S5b** depicts an example fit of a QENS signal on IN5 for a sample containing PbS nanoparticles and oleic acid in the d-hexane solvent, at $T$=239K, for one example $q$-value. No $q$-dependence was imposed on the fits for the IN5 QENS data. The resulting fit parameters are according to equation (2) and shown in **Figures S5c-f and Figure S6**. The fits for the two Lorentzian contributions seen on IN5 result in one narrow contribution at the limit of the range visible on IN16B (**Figure S5c**) and one broad contribution far beyond the range of IN16B (**Figure S5d**). The impact of the NCs on the solvent appears to be small within the accuracy of the data and fits at least at low $q$-values. The effect is more notable at high $q$, where, however, the associated intensities of the first Lorentzian decrease. Overall, this small effect of the presence of the NCs is probably due to the depletion of the oleic acid from the solvent in the presence of the NCs. In this picture, the solvent dynamics is slightly altered by the amount of free oleic acid.

Due to the coherent scattering from the solvent ($d_{14}$-hexane component), a maximum of the Lorentzian intensities can be observed near or slightly above $q^2$=2Å$^{-2}$ (**Figures S5c+d**). This peak appears to shift to higher $q$ at lower temperatures, as expected.

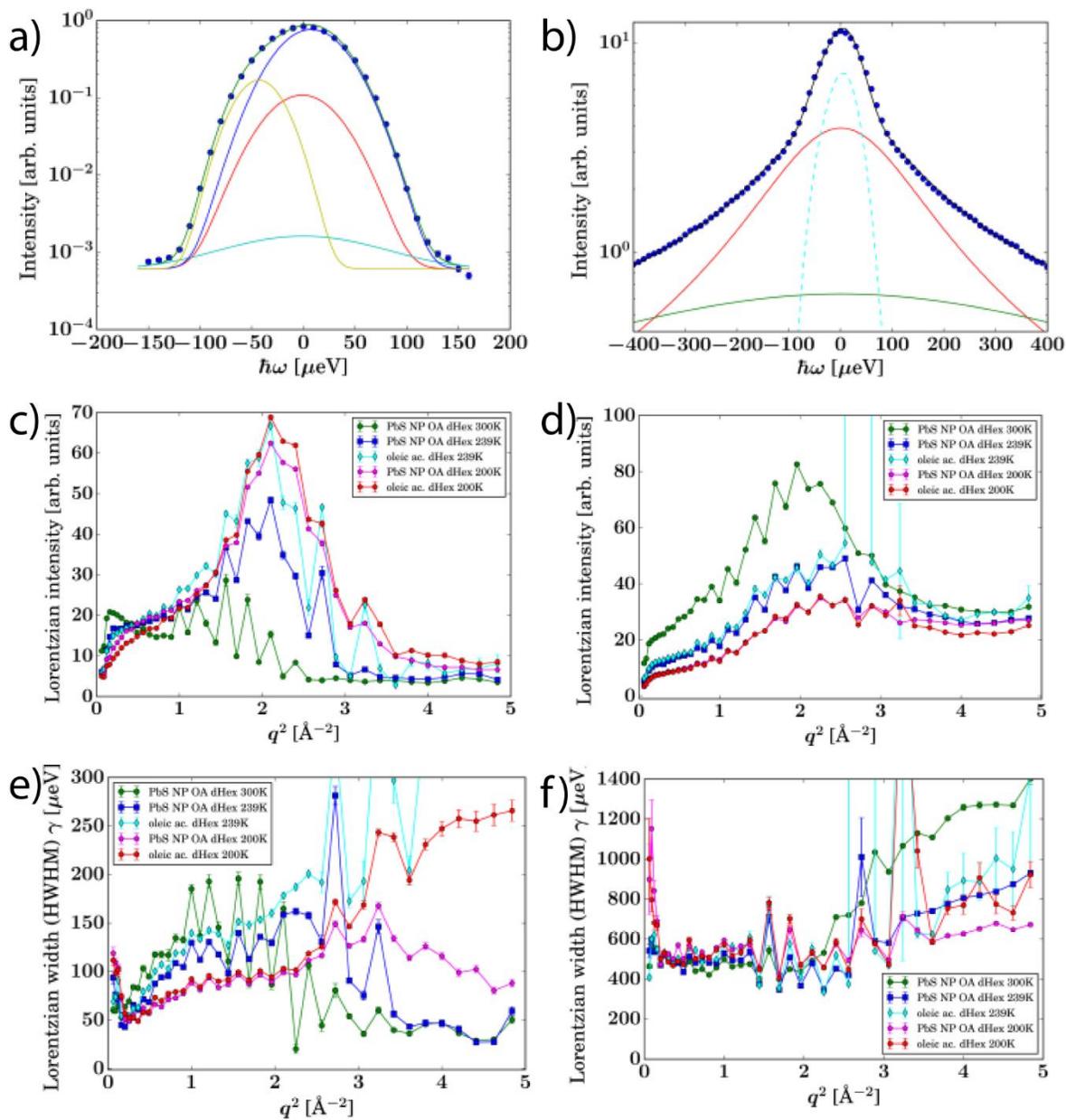

**Figure S5. a)** Energy resolution of IN5 measured using a Vanadium foil in the same geometry as the sample (symbols) and fit (green line superimposed on the symbols) consisting of a sum of four Gaussians (lines). Example for q=0.75Å$^{-1}$. **b)** Example spectrum (symbols) and fit (black line) of an IN5 QENS spectrum at q=1.0Å$^{-1}$ and T=239K of PbS NCs/OA/d$_{14}$-hexane. The red and green lines represent the Lorentzians $L_1$ and $L_2$ in equation (2), and the cyan dashed line denotes the apparent elastic ("Dirac") contribution. **c)** Intensity $I_1$ (q), **d)** Intensity $I_2$ (q), **e)** width $\sigma_1$ (q), and **f)** width $\sigma_2$ (q) of the two Lorentzians $L_1$ and $L_2$, which represent the solvent contributions.

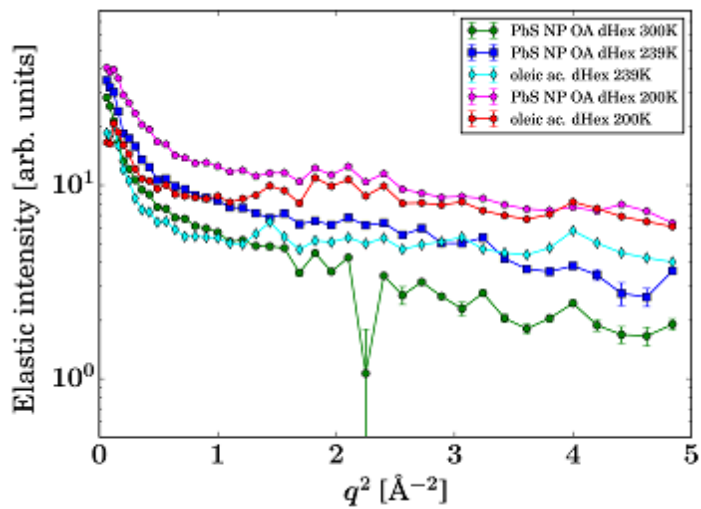

**Figure S6.** Apparent elastic contribution $I_3$ in equation (2) ("Dirac" δ-function) representing the immobile and slow QENS contributions much slower than the discernable motions within the energy resolution of IN5.

The apparent elastic contribution seen on IN5 (**Figure S6**) is consistent with the picture resulting from the IN16B fits: The elastic contribution is higher in the presence of the nanoparticles due to the immobile and slowly diffusing oleic acid bound to the NP. The elastic contribution decreases with rising temperature.